\documentclass[article,twocolumn,amsmath,floats,showpacs,nofootinbib]{revtex4}
\usepackage{graphicx}% Include figure files
\usepackage{textcomp}

\usepackage{hyperref}

\hypersetup{colorlinks=true}

\begin{document}

\title{Point-contact function of the electron–phonon interaction in zirconium}
\author{N. L. Bobrov, L. F. Rybalchenko, V. V. Fisun, and A. V. Khotkevich}
\affiliation{B.I.~Verkin Institute for Low Temperature Physics and
Engineering, of the National Academy of Sciences
of Ukraine, prospekt Nauki, 47, Kharkov 61103, Ukraine
Email address: bobrov@ilt.kharkov.ua}
\published {(\href{http://fntr.ilt.kharkov.ua/fnt/pdf/42/42-9/f42-1035r.pdf}{Fiz. Nizk. Temp.}, \textbf{42}, 1035 (2016)); ( \href{http://dx.doi.org/10.1063/1.4963330}{Low Temp. Phys.}, \textbf{42}, 811 (2016)}
\date{\today}

\begin{abstract}Experimental study of the electron–phonon interaction (EPI) spectra of zirconium was carried out
using Yanson's point-contact (PC) spectroscopy. The EPI spectral function and constant were
determined. Both homocontacts ($Zr-Zr$) and heterocontacts ($Zr-Cu$, $Zr-Ag$, and $Zr-Au$) were studied.
No contribution by copper, silver or gold to the heterocontact spectra was detected. The positions
of five phonon features of the EPI function were observed at energies 5-6, 9-11, 13-15,
18-20, and 27-29~$meV$. The observed diversity of the PC spectra was attributed to the anisotropy
of $Zr$.
\pacs{71.38.-k, 73.40.Jn, 74.25.Kc, 74.45.+c}
\end{abstract}
\maketitle

\section{Introduction}
The spectral functions (SF) of electron-phonon interaction
(EPI) (e.g., thermodynamic, \'{E}liashberg function, transport,
microcontact, etc.) are the most detailed characteristics
of the electron-phonon interaction: they define the renormalization
of the electron energy spectrum of metals as well as
many of the most important observable properties. The SFs
of EPI are related to the phonon density of states (PDS),
which makes their study also important from the perspective
of the crystal lattice dynamics.

To date, the experimental data on different kinds of the
SF of EPI in transition metals with a complex crystal lattice
are limited to point-contact (PC) spectroscopy data \cite{1,2} for $Co,
Tc, Re, Ru$, and $Os$. Attempts to reconstruct the pointcontact
EPI functions for group-IVb hcp metals - $Ti$, $Zr$ and
\emph{Hf} - were hindered by unfavorable physical and chemical
properties of these elements (rapid formation of a strong
oxide layer on the surface of each of the two electrodes
designed to create a pressed point contacts). In our studies
we used two different approaches to overcome these
technical difficulties. In the first approach, heterocontacts
$Zr$ - noble metal ($Cu, Ag, Au$) were created. In this case only
one surface of a $Zr$-electrode ends up in the region of the
point contact between the two electrodes. In addition, we
performed forming of these heterocontacts under a current
load \cite{3}. The second approach \cite{4} involved the use of break junctions \cite{5},
 when an elongated sample with a thinning (notch) in
the middle portion was first broken under liquid helium and
then a contact between clean non-oxidized surfaces was
established.

The aim of the present work is to obtain the pointcontact
EPI function and its numerical characteristics in $\alpha-Zr$
based on experimental data. Unfortunately, the utility of the
theoretical calculation of the SF of EPI for $Zr$ reported in
Ref. \cite{4} is very limited due to the fact that the calculated
graphs have been presented in the coordinates of $f(\omega/\omega_{max})$
($\omega_{max}$ is the maximum frequency of the phonon spectrum).

\section{Experimental methods}
The electrodes were prepared from a polycrystalline
material obtained by high-vacuum melting of zirconium
iodide using an electron beam gun. Polycrystalline copper,
silver and gold with the resistance ratio $R_{300K}/R_{4.2K}$ of about
several hundreds were also used. The electrodes measuring
$1.5\times 1.5\times 10~mm$ were spark-cut and chemically polished.
They were then washed in distilled water, dried and mounted
in the setup for forming microcontacts \cite{6}. The contacts were
formed using the shear method \cite{7} with electroforming as
described in detail in Ref. \cite{3}. The optimal point-contact resistance,
at which the spectra showed the maximum intensity
and the minimum background, was typically about several
tens of Ohms. The EPI spectra, which are proportional to the
second derivative of the current-voltage characteristics
(CVC), were measured using the modulation method at
acoustic frequencies at $4.2~K$.
\section{Processing of the measurement data}

The relation between the nonlinearities of the pointcontact
CVC in the normal state, which arise due to the
inelastic scattering of electrons in the contact region, and the
EPI spectrum are determined by the fundamental equation of
the Yanson theory of point-contact spectroscopy. Taking
into account the background term, it can be written as \cite{2}
\begin{equation}
\label{eq__1}
{\frac{{{d}^{2}}I}{d{{V}^{2}}}(V)=-\frac{4\pi{{e}^{3}}}{\hbar }n({{\varepsilon }_{F}}){{\Omega}_{\text{eff}}}\langle K \rangle {{ [{{g}_{pc}}(\omega)+B(\omega)]|}_{\omega=eV/\hbar}}.}
\end{equation}

Here, $d^{2}I/dV^{2}(V)$ is the voltage dependence of the second
derivative of the contact CVC, $g_{pc}(\omega)$ is the point-contact
EPI function, $B(\omega)$ is the background function, $n(\varepsilon_{F})$ is the
density of electron states at the Fermi level for a specific
spin direction, \emph{e} is the elementary charge, $\Omega_{\text{eff}}$ is the effective
volume of the phonon generation in the contact, $K$ is the
unnormalized point-contact form factor, and the angle brackets
denote averaging over the entire Fermi surface (FS). The
presence of the background function reflects the fact that the
second derivative of the CVC is non-zero for $eV>\hbar \Omega_{max}$,
while the EPI function should be identically zero beyond the
phonon spectrum edge. The quantities $\Omega_{\text{eff}}$ and $\langle K \rangle$ depend
on the geometry and cleanliness of the contact, and in addition,
$K$ also depends on the orientation of the contact axis
with respect to the crystallographic directions in the metal.
Equation \eqref{eq__1} can be expressed through the experimentally measurable characteristics in the form suitable for reconstructing
the EPI function $g_{pc}(\omega)$ \cite{8}:

\begin{equation}
\label{eq__2}
{{g}_{pc}}(\omega)+B(\omega)=C{{\tilde{V}}_{2}}(V)V_{1,0}^{-2}{{R_{0}^{1/2}|}_{V=\hbar \omega /e}}.
\end{equation}
Here, $\tilde{V}_{2}(V)\propto d^{2}I/dV^{2}(V)$ and $V_{1,0}\sim dV/dI$ and $R_{0}$ are the
modulating voltage $V_{1}(V)$ and the resistance of the point contact
$R$ for $V\rightarrow 0$. The dimensional constant $C$ is defined for
the model of a clean round aperture in the approximation of
the isotropic quadratic dispersion law of electrons as

\begin{equation}
\label{eq__3}
C=-\frac{3}{8}{{\left( \frac{\hbar }{2\pi } \right)}^{1/2}}{{k}_{F}}{{v}_{F}},
\end{equation}
where $k_F$ and $v_F$ are the Fermi wave vector and the Fermi
velocity, respectively. For $Zr$, the values of $k_F$ and $v_F$ were
calculated from low-temperature lattice constants \cite{9} $c/a=5.141~\AA/3.229~\AA$
using the relation \cite{10} $k_{F}={(3\pi^{2}Z/\Omega)}^{1/3}$,
where $Z$ is the number of conduction electrons per unit cell
and $\Omega$ is the volume of the unit cell (for hcp lattice $\Omega=(\sqrt{3}/2)a^{2}c$).
 Using $Z=1.2$ for $Zr$ \cite{11},  we obtain $k_{F}=0.915\times 10^{-8}~cm^{-1}$
and $v_{F}=1.06\times10^{8}~cm/s$.

Since in the experiment the voltage at the second harmonic
of the modulating current was measured $V_{2}(V)\propto dV^{2}(V)/d^{2}I$, this dependence is recalculated into
$\tilde{V}_{2}(V)\propto d^{2}I(V)/dV^2$, according to the theorem about the
derivative of an inverse function. The background function
in Eq. \eqref{eq__1} was chosen as in Ref. \cite{12}. When assessing the
effective contact diameter from the resistance, it is convenient
to use the Sharvin equation for the case of spherical FS
and ballistic regime, written in the form

\begin{equation}
\label{eq__4}
d=\frac{4}{e{{k}_{F}}}{{\left( \frac{\pi \hbar }{{{R}_{0}}} \right)}^{1/2}}.
\end{equation}

\section{Experimental results}

We studied both homocontacts $Zr-Zr$ and heterocontacts
$Zr-Cu$, $Zr-Ag$, and $Zr-Au$. Despite the fact that the interpretation
of the homocontact spectra is much simpler and does
not involve many assumptions required for the processing of
the heterocontact spectra, the use of the latter for reconstructing
the EPI function in zirconium was completely
justified. As for \emph{Ta} contacts \cite{3} the contribution of the noble
metals is not observable in the heterocontact spectra. This is
due not only to their lower intensity, but also the peculiarities
of the summation of the partial contributions into the
resulting spectrum of a heterocontact.

The heterocontact spectrum does not coincide with the
half-sum of the contributions from the homocontact spectra
of the same diameter. This is due to the fact that
the point-contact form factors of metals with different
momenta in the heterocontact are not equal to each other
and therefore do not match the form factor of the homocontact.
In a metal with a greater $p_F$, the relative phase volume
of nonequilibrium occupied states is lower due to
the reflection of electron trajectories from the interface,
resulting in a relative increase of the inelastic relaxation
time $\tau_{\varepsilon}$, while the intensity of the PC spectrum is proportional
to $d/v_{F}\tau_{\varepsilon}$.

In other words, in a heterocontact, the contribution of
a metal with a lower Fermi momentum to the resultant spectrum
will be greater than the half of the spectrum corresponding
to a homocontact of the same diameter; conversely, for a
metal with a large momentum, it is, respectively, less.

It should be also noted that the PC spectrum of a heterocontact
is symmetrical with respect to the polarity of the
applied voltage, and at $T=0$ can be represented as the sum
of the PC spectra of each of the contacting metals.

The intensity ratio of the PC spectra of metals forming
the heterocontact does not depend on the elastic scattering
path of electrons. Introducing impurities in one metal also
reduces the partial contribution to the PC spectrum of the
clean one \cite{13}.
\begin{figure}[]
\includegraphics[width=8.5cm,angle=0]{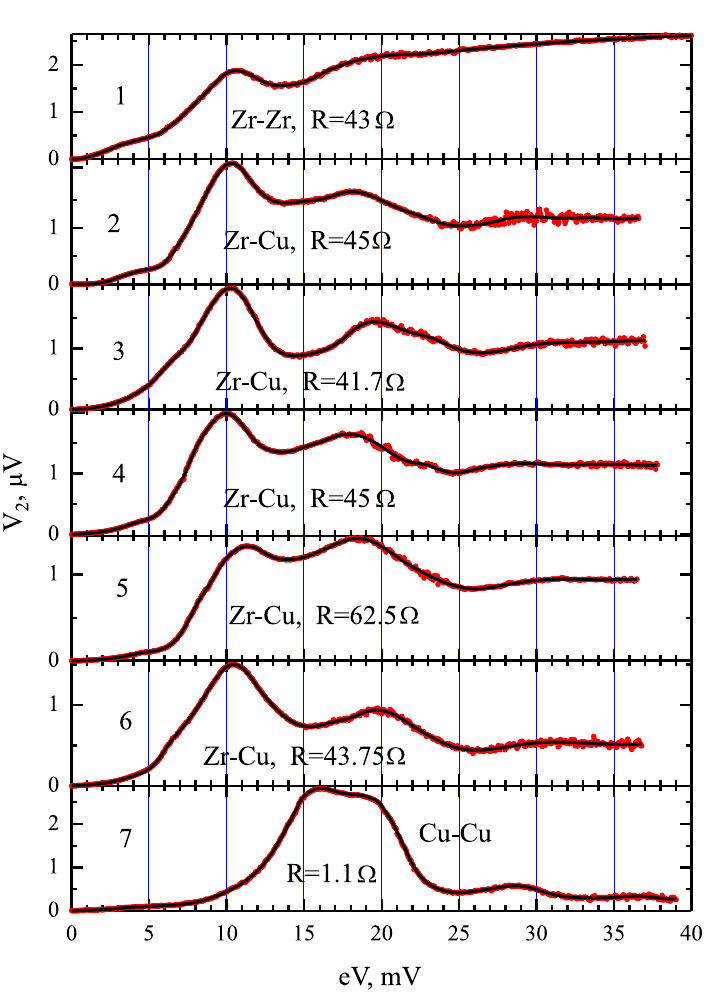}
\caption[]{Point-contact EPI spectra $V_{2}\sim d^{2}V/dI^{2}$, T=4.2~$K$: $V_{1}(0)=0.55$(1);
0.89(2); 0.763(3); 0.878(4); 1.028(5); 1.196(6); and 0.655(7)~$mV$.}
\label{Fig1}
\end{figure}

\begin{figure}[]
\includegraphics[width=8.5cm,angle=0]{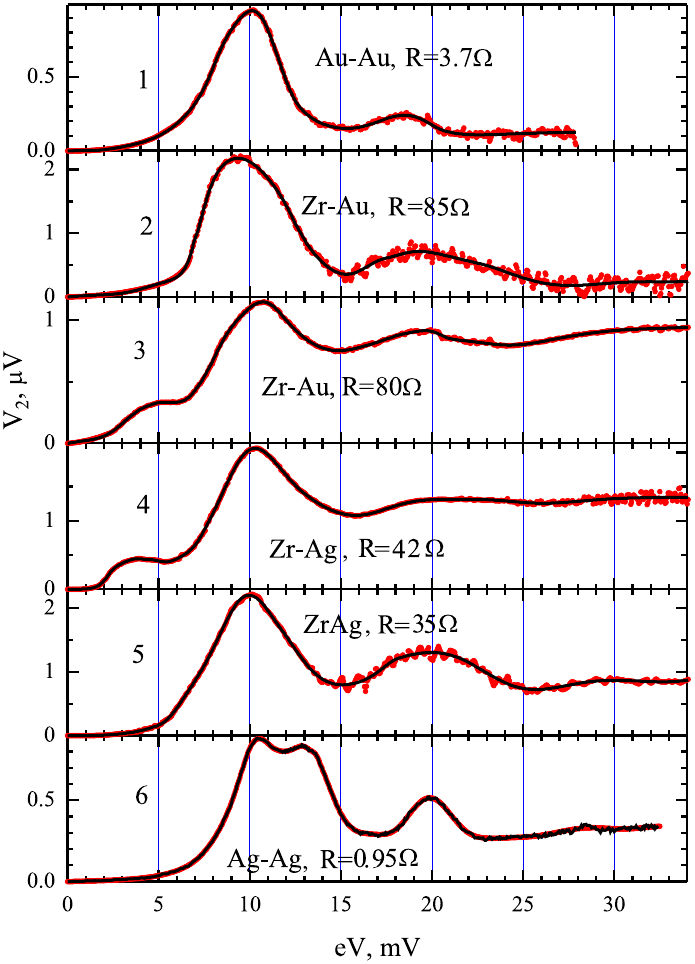}
\caption[]{Point-contact EPI spectra $V_{2}\sim d^{2}V/dI^{2}$, T=4.2~$K$ (except curve 6):
$V_{1}(0)=0.55$(1); 1.213(2); 1.355(3); 0.778(4); 0.933(5)~$mV$; curve 6 was
acquired at $V_{1}(0)=0.45~mV$, $T=1.6~K$.}
\label{Fig2}
\end{figure}

Let us estimate the expected contribution of gold to the
spectrum of a $Zr-Au$ heterocontact in the free-electron
approximation (due to similar Fermi parameters for \emph{Ag} and
\emph{Cu}, the result for this pair with \emph{Zr} should be similar).

As follows from Ref. \cite{13}, the intensity ratio of the partial
contributions to the heterocontact spectrum is
\begin{equation}
\label{eq__5}
L={{{\left( \frac{{{d}^{2}}I}{d{{V}^{2}}} \right)}_{1}}}\Big/{{{\left( \frac{{{d}^{2}}I}{d{{V}^{2}}} \right)}_{2}}}=\frac{{{v}_{{{F}_{2}}}}}{{{v}_{{{F}_{1}}}}}{{\left( \frac{{{p}_{{{F}_{2}}}}}{{{p}_{{{F}_{1}}}}} \right)}^{2}}\frac{g_{pc}^{(1)}}{g_{pc}^{(2)}}.
\end{equation}
Since a similar relation has been obtained in Ref. \cite{13} for
model metals with the same shape and absolute intensity of
the PC function of EPI but with different Fermi velocities
and momenta, the factor $g_{pc}^{(1)}/g_{pc}^{(2)}$
 is absent in Ref. \cite{13}.
\begin{figure}[]
\includegraphics[width=8.5cm,angle=0]{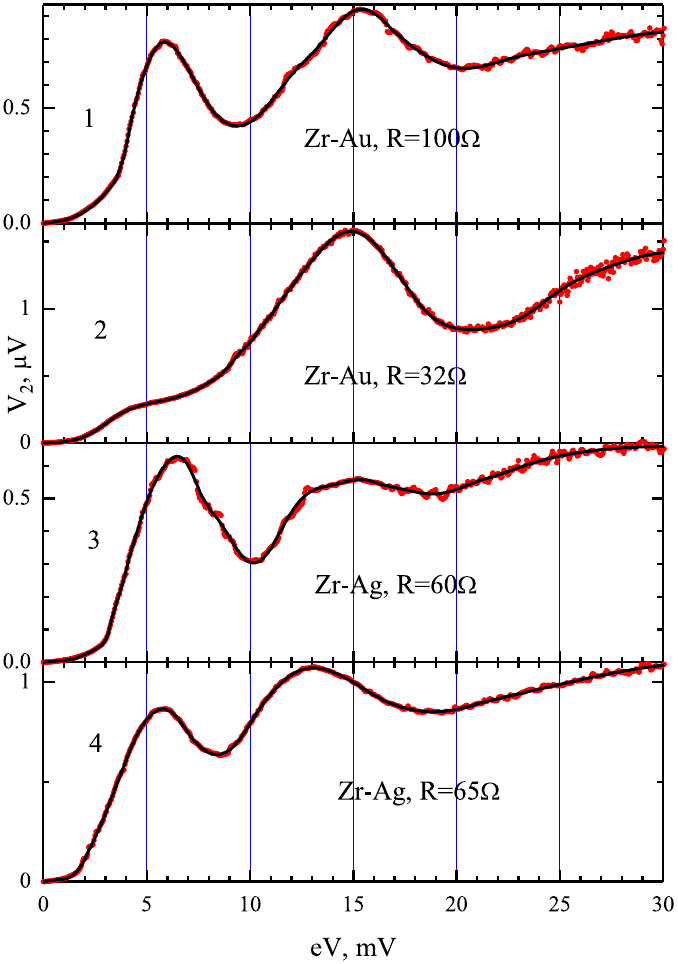}
\caption[]{Point-contact EPI spectra $V_{2}\sim d^{2}V/dI^{2}$, T=4.2~$K$: $V_{1}(0)=0.516$(1);
 1.633(2); 1.646(3); and 1.648(4)~$mV$.}
\label{Fig3}
\end{figure}

\begin{figure}[]
\includegraphics[width=8.5cm,angle=0]{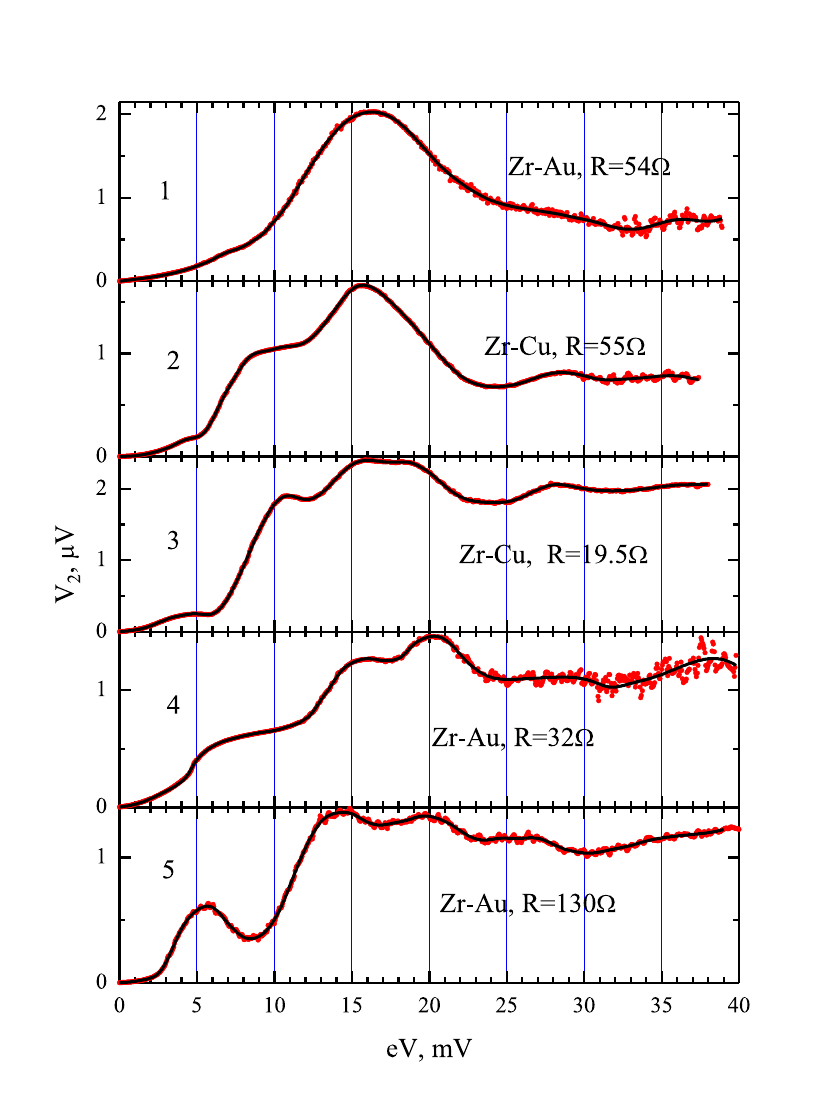}
\caption[]{Point-contact EPI spectra $V_{2}\sim d^{2}V/dI^{2}$, T=4.2~$K$:
$V_{1}(0)=0.557$(1); 1.062(2); 0.751(3); 1.06(4); and 1.777(5)~$mV$.}
\label{Fig4}
\end{figure}

\begin{figure}[]
\includegraphics[width=8.5cm,angle=0]{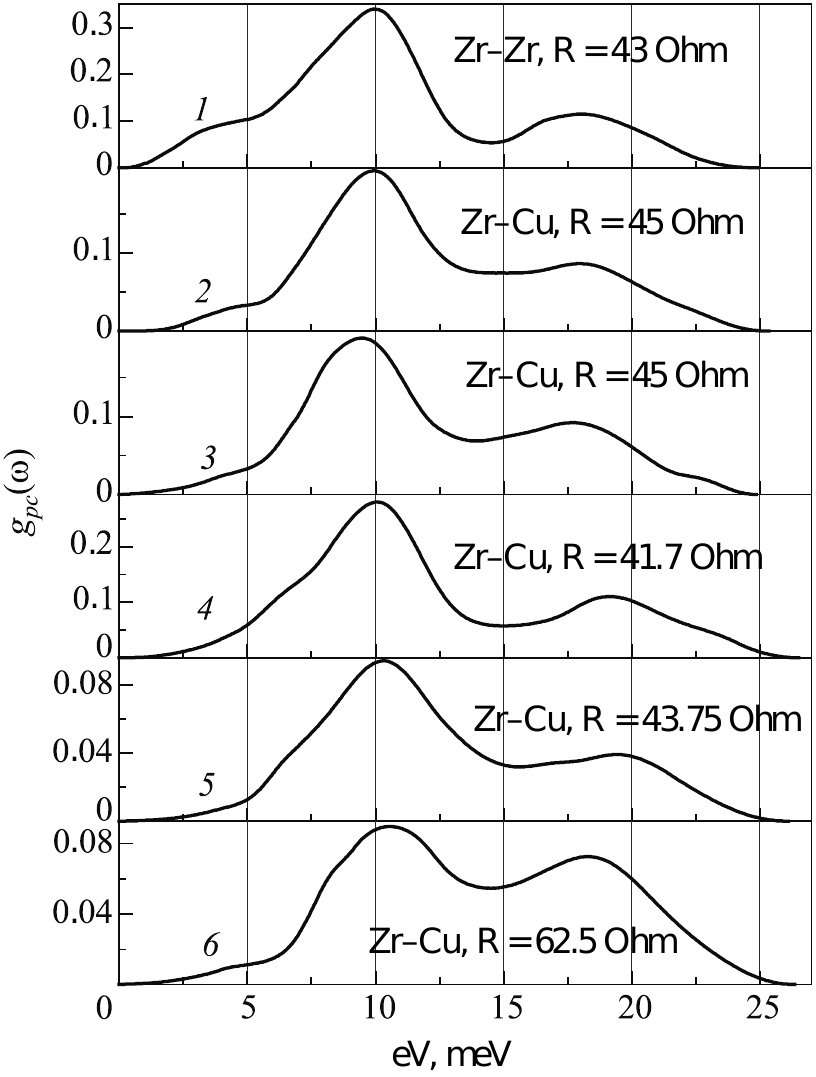}
\caption[]{Point-contact EPI function reconstructed from the spectra in Fig. \ref{Fig1}:
$\lambda$=0.66(1); 0.34(2); 0.36(3); 0.46(4); 0.16(5); and 0.18(6).}
\label{Fig5}
\end{figure}

\begin{figure}[]
\includegraphics[width=8.5cm,angle=0]{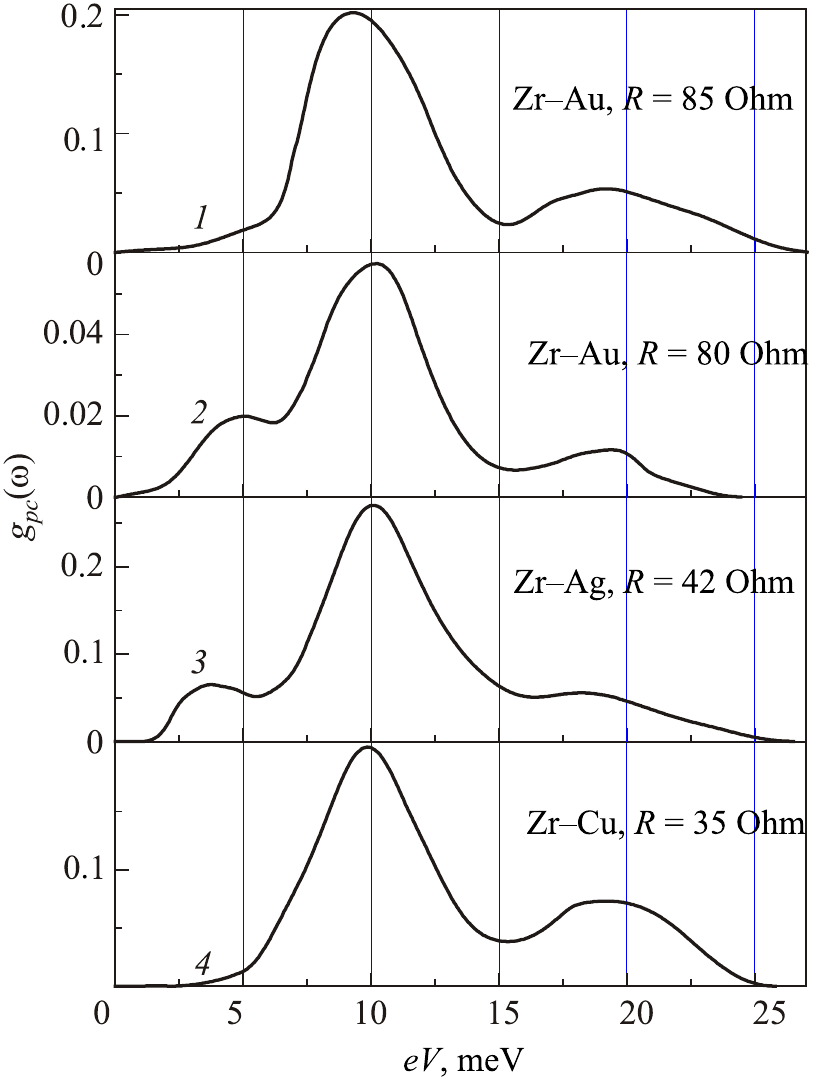}
\caption[]{Point-contact EPI function reconstructed from the spectra in Fig. \ref{Fig2}:
$\lambda$=0.295(1); 0.103(2); 0.45(3); and 0.285(4).}
\label{Fig6}
\end{figure}

\begin{figure}[]
\includegraphics[width=8.5cm,angle=0]{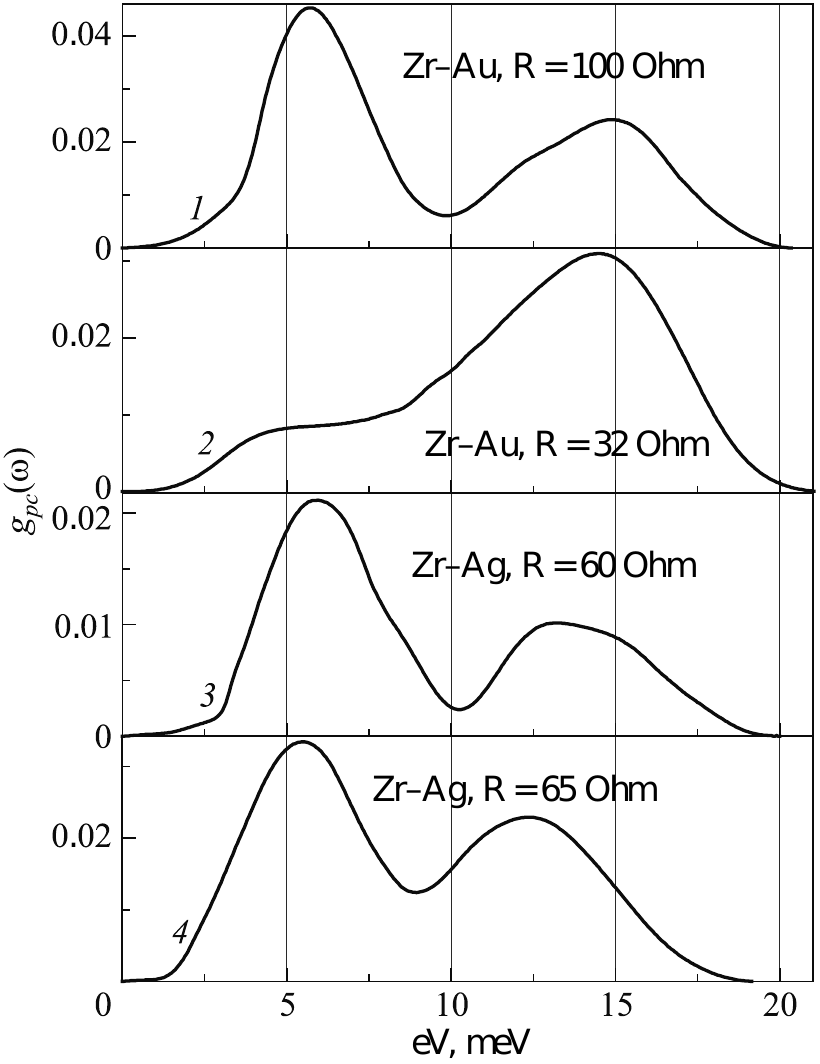}
\caption[]{Point-contact EPI function reconstructed from the spectra in Fig. \ref{Fig3}:
$\lambda$=0.085(1); 0.053(2); 0.04(3); and 0.084(4).}
\label{Fig7}
\end{figure}

\begin{figure}[]
\includegraphics[width=8.5cm,angle=0]{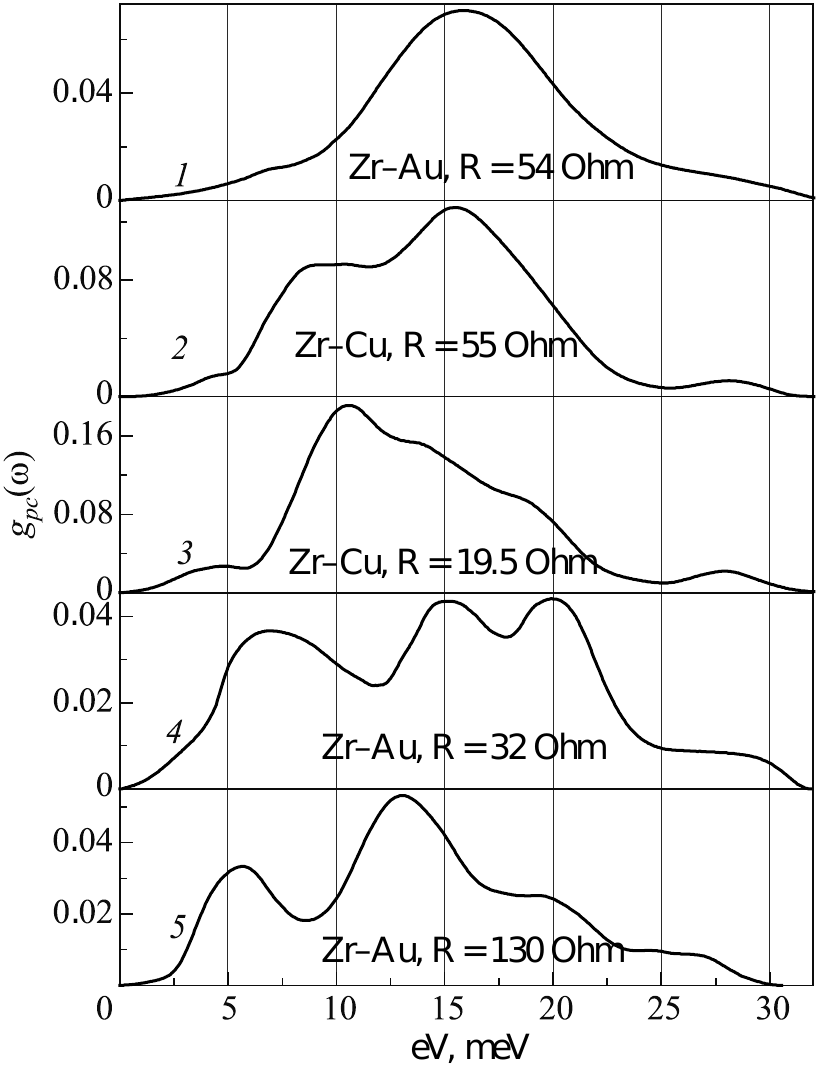}
\caption[]{Point-contact EPI function reconstructed from the spectra in Fig. \ref{Fig4}:
$\lambda$=0.115(1); 0.254(2); 0.367(3); 0.138(4); and $\lambda=0.124$(5).}
\label{Fig8}
\end{figure}

\begin{figure}[]
\includegraphics[width=8.5cm,angle=0]{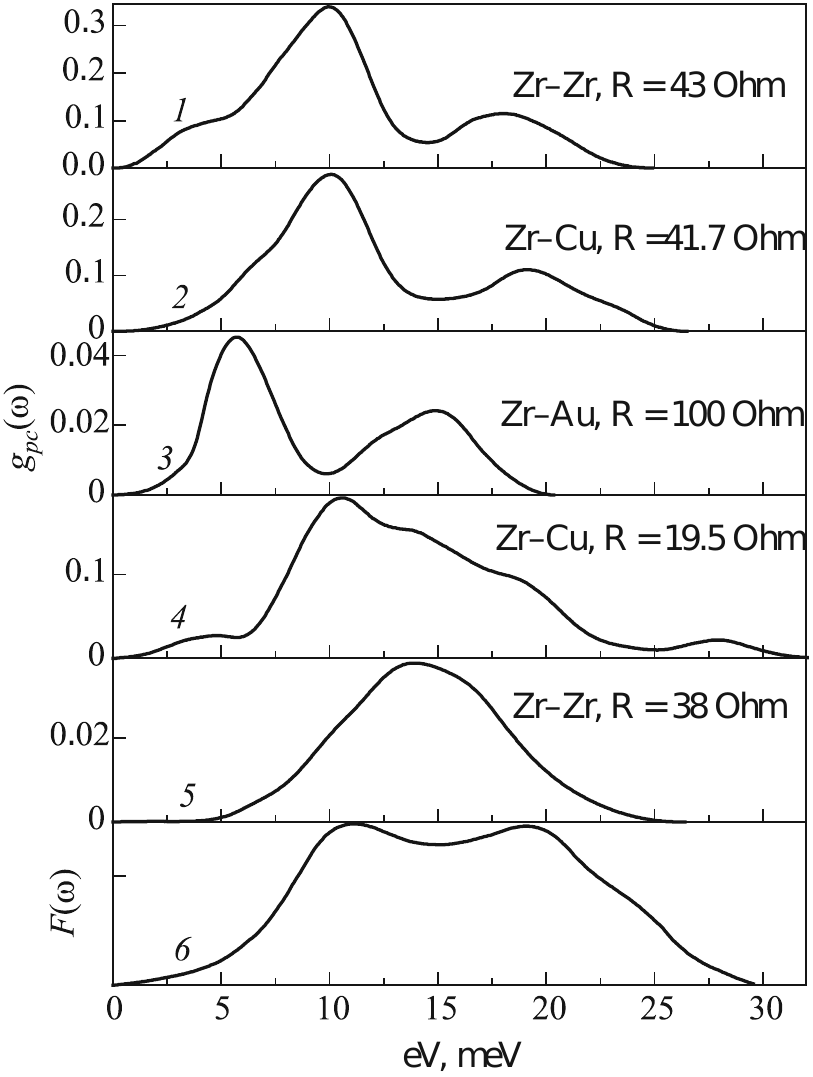}
\caption[]{Curves 1-4 are the point-contact EPI functions, selected from Figs. \ref{Fig5}-\ref{Fig8}, which presumably correspond to different crystallographic directions:
$\lambda$=0.66(1), 0.46(2), 0.085(3), and 0.367(4). Curve 5 shows the EPI function
from Ref. \ref{Fig4} with the contact axis aligned to the crystallographic $c$-axis
of \emph{Zr}, $\lambda$=0.055. Curve 6 shows the function of the phonon density of states
(data from neutron measurements \cite{14}).}
\label{Fig9}
\end{figure}

The absolute intensity of the PC function of EPI (Fig. \ref{Fig3})
that we obtained for zirconium homocontacts in the freeelectron
approximation for $g_{pc max}^{Zr}=0.34$, $g_{pc max}^{Au}=0.079$ \cite{2},
$v_{F}^{Zr}=1.06\times 10^{8}~cm/s$, $v_{F}^{Au}=1.4\times 10^{8}~cm/s$,
$k_{F}^{Zr}=0.915\times 10^{-8}~cm^{-1}$, $k_{F}^{Au}=1.21\times 10^{-8}~cm^{-1}$
 yields the contribution
of gold to the spectrum of the heterocontact $L_{Au}\simeq 0.1$. Note
that this is the upper-bound estimate. While sufficiently large
statistics has been accumulated for gold and the intensity of
its EPI function has been determined with high confidence, it
is not the case for \emph{Zr}. As follows from Fig. \ref{Fig1}, the spectrum
from which the EPI function of zirconium was restored is
characterized by a high level of background, which is not
typical for ballistic point contacts. Thus, it is likely that an
improved technology and/or the use of high-quality single
crystals for the electrodes can enable obtaining more intense
spectra. The importance of technology can be seen by comparing
the value of the EPI function obtained in the previous
study  \cite{4} using the break-junction technique with the present
data. Despite the use of high-quality single crystal \emph{Zr} for the
electrodes in Ref. \cite{4}, the intensity of the EPI functions they
have obtained is 9-fold lower (cf. curves 1 and 5 in Fig. \ref{Fig9}).

The main advantage of heterocontacts is significantly
higher quality of their spectra. First, the background level is
several times smaller (see Fig. \cite{1}). Second, the phonon structure
is much better visible, especially in the high-frequency
region. A higher yield of high-quality spectra compared to
homocontacts should also be noted. Thus, for determining as
accurately as possible the EPI function for many transition
metals, heterocontacts with noble metal represent a preferred
alternative to homocontacts.

All the obtained spectra may be divided into three
groups. The most numerous group is shown in Figs. \ref{Fig1} and \ref{Fig2}.
As can be seen in these figures, there are two major maxima
present in the spectra in the regions 9-11 and 18-20~$meV$. In
addition, many spectra also exhibit a soft mode around
$4-6~meV$. These figures also show the spectra of homocontacts
of \emph{Cu}, \emph{Ag}, and \emph{Au} for comparison. Despite the fact that
some phonon features in the spectra of noble metals and \emph{Zr}
heterocontacts are sufficiently close in energy, the overall
shape of the heterocontact spectra is markedly different from
those of noble metals. The spectrum edge for this group is in
the vicinity of $25~meV$.

The second group of spectra is shown in Fig. \ref{Fig3}. As in
the previous case, there are two major maxima present in the
spectra but at energies of 5-6 and 13-15~\emph{meV}. There are no
features in the spectra that could match the energies of those
in \emph{Ag} and \emph{Au}, forming the heterocontacts. The spectrum
edge for this group is in the vicinity of 20~\emph{meV}.

The third group is presented in Fig. \ref{Fig4}. The positions of phonon
features and the shape of the spectra vary widely in this
group, covering the previous cases. The hallmark of this group
is the presence of a high-frequency maximum at 27-29~$meV$.
The phonon spectrum edge is located at about 31~$meV$.

Figures \ref{Fig5}, \ref{Fig6}, \ref{Fig7}, and \ref{Fig8} display the EPI functions reconstructed
from the graphs in Figs. \ref{Fig1}, \ref{Fig2}, \ref{Fig3}, and \ref{Fig4}, respectively.
All the values of $g_{pc}$ and $\lambda$ reported in the captions
to Figs. \ref{Fig5}-\ref{Fig9} for the cases of heterocontacts were calculated
in the same way as for the $Zr-Zr$ homocontacts of the same
resistance.

In order to determine $g_{pc}$ and $\lambda$ in a \emph{Zr}-heterocontact
more accurately, a correction factor can be estimated:

\begin{equation}
\label{eq__6}
{{K}_{\text{corr}}}=\frac{\lambda _{pc}^{\text{get}}}{\lambda _{pc}^{\text{gom}}}=\frac{g_{pc}^{\text{get}}}{g_{pc}^{\text{gom}}}=\frac{2\left\langle {{K}_{Zr}} \right\rangle {{d}_{Zr}}}{\left\langle {{K}_{0}} \right\rangle {{d}_{\text{get}}}}.
\end{equation}

Here $\langle K_{0}\rangle$ and $\langle K_{Zr}\rangle$ are the form factors for homo- and heterocontacts
averaged over the Fermi surface and $d_{Zr}$ and $d_{get}$
are the diameters of homo- and heterocontacts, respectively.
The factor 2 in the numerator is introduced to account for
two-fold lower phonon generation in \emph{Zr} heterocontact compared
to the homocontact of the same diameter.

When calculating the diameter of the heterocontact on
the zirconium side, we use the expression \cite{3}
\begin{equation}
\label{eq__7}
{{d}_{\text{get}}}={{d}_{Zr}}{{\left[ 2{{\left\langle {{\alpha }_{1}}D\left( {{\alpha }_{1}} \right) \right\rangle }_{{{V}_{Z}}>0}} \right]}^{-1/2}}.
\end{equation}

Let us denote $p_{F_{1}}/p_{F_{2}}=b$ and $v_{F_{1}}/v_{F_{2}}=c$ and assume
$b,c<1$. Then
\begin{equation}
\label{eq__8}
D({{\alpha }_{1}})=\frac{4b{{\alpha }_{1}}\sqrt{\alpha _{1}^{2}+{{b}^{-2}}-1}}{c{{\left( {{\alpha }_{1}}+\frac{b}{c}\sqrt{\alpha _{1}^{2}+{{b}^{-2}}-1} \right)}^{2}}}\ ,
\end{equation}

\begin{equation}
\label{eq__9}
2{{\left\langle {{\alpha }_{1}}D({{\alpha }_{1}}) \right\rangle }_{{{V}_{Z}}>0}}=2\int\limits_{0}^{1}{{{\alpha }_{1}}}D({{\alpha }_{1}})d{{\alpha }_{1}}\ .
\end{equation}

Now we write the expression for the form factors of a
clean heterocontact \cite{13}
\begin{equation}
\label{eq__10}
{{K}_{s}}\left( \mathbf{p},\mathbf{{p}'} \right)=\frac{D({{\mathbf{p}}_{s}}=\mathbf{p})D({{\mathbf{p}}_{s}}=\mathbf{{p}'})}{4{{\left\langle \alpha D(\alpha ) \right\rangle }_{s}}}{{K}_{0}}\left( \mathbf{p},\mathbf{{p}'} \right)\ ,
\end{equation}
where $K_{0}(\mathbf{p},\mathbf{{p}'})$ are the form factors of clean homocontacts,
same for both metals, and $K_{s}(\mathbf{p},\mathbf{{p}'})$ are the form factors of
zirconium and the noble metal.

To determine the average form factor of zirconium heterocontacts
in ballistic mode, it is first necessary to calculate
the numerator of Eq. \eqref{eq__10}
\[{{A}_{s=Zr}}={{\left\langle {{D}_{12}}({{\mathbf{p}}_{s}}=\mathbf{p}){{D}_{12}}({{\mathbf{p}}_{s}}=\mathbf{{p}'}){{K}_{0s}}({{\mathbf{p}}_{s}},\mathbf{{p}'}) \right\rangle }_{{{V}_{z}}>0}}\]

In the aperture model \cite{13}
\begin{equation}
\label{eq__11}
{{K}_{0s}}={{K}_{0}}\left( \mathbf{p},\mathbf{{p}'} \right)=\frac{\left| {{v}_{z}}{{{{v}'}}_{z}} \right|\theta (-{{v}_{z}}{v}')}{\left| {{v}_{z}}\mathbf{{v}'}-{{{{v}'}}_{z}}\mathbf{v} \right|}\ ,
\end{equation}
where $\theta$ is the Heaviside function.

For a spherical Fermi surface, after rearrangement we
obtain \cite{3}
\begin{equation}
\label{eq__12}
A=16\int\limits_{0}^{1}{dx\int\limits_{0}^{1}{dy}}\left. \frac{xD({{\alpha }_{1}}=x)D({{\alpha }_{1}}=y)}{{{\left( m+n \right)}^{2}}} \right|,\quad y>x,
\end{equation}
\[n={{\left( 1-{{x}^{2}} \right)}^{1/4}}+{{\left( 1-{{x}^{2}}/{{y}^{2}} \right)}^{1/4}}\quad ,\]

\[m=\left\{ 8{{\left[ (1-{{x}^{2}})\left( 1-\frac{{{x}^{2}}}{{{y}^{2}}} \right) \right]}^{1/4}}\times  \right.\]
\[\times {{\left. \left[ {{(1-{{x}^{2}})}^{1/2}}+{{\left( 1-\frac{{{x}^{2}}}{{{y}^{2}}} \right)}^{1/2}} \right] \right\}}^{1/4}}\ .\]

Assuming $b=c=0.673$ for the couple $Zr-Cu$, we obtain
$A=0.19$ and ${{\left[ 2{{\left\langle {{\alpha }_{1}}D({{\alpha }_{1}}) \right\rangle }_{{{V}_{Z}}>0}} \right]}^{-1/2}}=0.86$, hence $K_{\text{corr}}^{Cu}=1.77$
For $Zr-Ag$ and $Zr-Au$ heterocontacts \emph{b} and \emph{c} are
approximately equal. Assuming $b=c=0.76$, we obtain
$A=0.21$ and ${{\left[ 2{{\left\langle {{\alpha }_{1}}D({{\alpha }_{1}}) \right\rangle }_{{{V}_{Z}}>0}} \right]}^{-1/2}}=0.9$, then $K_{\text{corr}}^{Ag,Au}=1.87$.

Thus, given the correction coefficients, the maximum values
for the $Zr-Cu$ heterocontacts are: $\lambda=0.81$, $g_{pc}^{\max }=0.5$
(Fig. \ref{Fig5}, panel 4 and Fig. \ref{Fig9}, panel 2; $R=41.7~\Omega$). Accordingly,
for $Zr-Ag$ we have: $\lambda=0.84$, $g_{pc}^{\max }=0.51$ (Fig. \ref{Fig6}, panel 3;
$R=42~\Omega$). Finally, for $Zr-Au$: $\lambda=0.55$, $g_{pc}^{\max }=0.38$
(Fig. \ref{Fig6},
panel 1; $R=85~\Omega$). Recall that for the $Zr-Zr$ homocontact we
obtained: $\lambda=0.66$, $g_{pc}^{\max }=0.34$
(Fig. \ref{Fig5}, panel 1 and Fig. \ref{Fig9},
panel 1; $R=43~\Omega$).

The fact that the refined values for the $Zr-Cu$ and $Zr-Ag$
heterocontacts turned out to be very close to each other indicate
that the technological limits are reached and these contacts
are apparently in the ballistic limit. Somewhat lower
values for the couple $Zr-Au$ reflect the fact that a smaller
number of these spectra were considered.

Thus, the use of heterocontacts allowed not only to refine
the position of the phonon features in the spectra, but also to
refine the value of the constant $\lambda$ and the EPI function $g_{pc}$.

Note that this refinement factor, in addition to a spherical
Fermi surface, assumes a geometrically symmetric heterocontact
with a flat (mirror) boundary between the two sides.
Therefore, any numerical values obtained with these coefficients
are just estimates.

Finally, Fig. \ref{Fig9} shows all of the above types of pointcontact
EPI functions grouped together for clarity; they are
apparently related with different orientations of the contact
axis with respect to the crystallographic directions in the \emph{Zr}
electrode. The figure also shows the EPI function reconstructed
in Ref. \cite{4}. For the latter, it is known that the direction of the
point-contact axis was aligned to the \emph{c}-axis of the single crystal
of \emph{Zr}. Note that the EPI function (curve 1 in Fig. \ref{Fig8}) virtually
matches its shape indicating the same orientation of the
contact axis. It is interesting that curve 4 in Fig. \ref{Fig9} includes all
the five features occurring in different types of spectra. The
last panel in Fig. \ref{Fig9} shows the function of the phonon density of
states (curve 6). Unfortunately, as polycrystalline electrodes
were used in the experiments, it is not possible to determine
the direction of the point-contact axes. However, since most of
the spectra exhibit features in the regions 9-11 and
18-20~$meV$, which coincides with the function of the phonon
density of states, we can assume that these phonon modes
dominate for the majority of the crystallographic directions.

\section{Summary}

\begin{enumerate}
\item {Both $Zr-Zr$ homocontacts and heterocontacts of \emph{Zr} and
noble metals, \emph{Cu}, \emph{Ag}, and \emph{Au}, were employed for the experimental study of the EPI in $\alpha$-zirconium using
Yanson's point-contact spectroscopy.}
\item {It was found that the contribution of noble metals to the
PC spectra of $Zr-Cu$, $Zr-Ag$, $Zr-Au$ heterocontacts (second
derivative of the CVCs) is not observable. In comparison
with the characteristics of \emph{Zr} homocontacts, the
heterocontact spectra have a lower background and
sharper high-frequency features.}
\item {From the experimental data, the graphs of the pointcontact
EPI function were reconstructed, its absolute values
were determined, and the value of the EPI constant
was estimated. Manifestations of the anisotropy of the
EPI spectrum in zirconium were observed.}

\end{enumerate}
The work was supported by the National Academy of
Sciences of Ukraine within the Project No. FZ 3-19. The
authors are grateful to Yu. G. Naydyuk for his comments
and suggestions made during the discussion of the paper.


\begin{thebibliography}{}

\bibitem{1} I. K. Yanson, Zh. Eksp. Teor. Fiz. \textbf{66}, 1035 (1974) [\href{http://www.jetp.ac.ru/cgi-bin/dn/e_039_03_0506.pdf}{Sov. Phys. JETP} 39,
506 (1974)].
\bibitem{2}A. V. Khotkevich and I. K. Yanson, \href{http://link.springer.com/book/10.1007%2F978-1-4615-2265-2}{Atlas of Point-Contact Spectra of
Electron-Phonon Interactions in Metals} (Kluwer Academic Publishers,
Boston/Dordrecht/London, 1995).
\bibitem{3}N. L. Bobrov, L. F. Rybal'chenko, V. V. Fisun, and I. K. Yanson, \href{http://fntr.ilt.kharkov.ua/fnt/pdf/13/13-6/f13-0611r.pdf}{Fiz.
Nizk. Temp.} \textbf{13}, 611 (1987) [Sov. J. Low Temp. Phys. \textbf{13}, 344
(1987)]; \href{https://arxiv.org/pdf/1512.01800.pdf}{ arXiv:1512.01800}.
\bibitem{4}V. V. Khotkevich, A. V. Khotkevich, A. P. Zhernov, T. N. Kulagina,
and E. K. Fol'k, Bull. Khark. Natl. Karazin Univ. Ser. Phys. \textbf{476}, 96
(2000)
\bibitem{5} J. Moreland and J. W. Ekin, \href{http://scitation.aip.org/content/aip/journal/jap/58/10/10.1063/1.335608}{J. Appl. Phys.} \textbf{58}, 3888 (1985).
\bibitem{6}N. L. Bobrov, L. F. Rybal'chenko, A. V. Khotkevich, P. N. Chubov, and I.
K. Yanson, Inventor's Certificate 1631626 USSR, M. Kl. 5 N 01. L 21/28
(1991), Bul. no. 8.
\bibitem{7}P. N. Chubov, A. I. Akimenko, and I. K. Yanson, Inventor's Certificate
834803 USSR, M.Kl. 3 N. 01 L 21/28. (1981), Nul. No. 20.
\bibitem{8}A. V. Khotkevich, Ph.D. dissertation, Physical and Mathematical
Sciences, FTINT, Kharkov (1990).
\bibitem{9}V. A. Finkel, Low-Temperature x-Ray Diffraction in Metals (Metallurgy,
Moscow, 1971) [in Russian].
\bibitem{10}W. A. Harrison, \href{https://www.abebooks.co.uk/book-search/title/solid-state-theory/author/harrison-walter/}{Solid State Theory} (McGraw Hill, New York, 1970).
\bibitem{11}G. Bambakidis, \href{http://onlinelibrary.wiley.com/wol1/doi/10.1002/pssb.2220540150/abstract}{Phys. Status Solidi (b)} \textbf{54}, K57 (1972).
\bibitem{12}I. K. Yanson, I. O. Kulik, and A. G. Batrak, \href{http://link.springer.com/article/10.1007/BF00117430}{J. Low Temp. Phys.} \textbf{42}, 527
(1981).
\bibitem{13}R. I. Shekhter and I. O. Kulik, \href{http://fntr.ilt.kharkov.ua/fnt/pdf/9/9-1/f09-0046r.pdf}{Fiz. Nizk. Temp.} \textbf{9}, 46 (1983) [Sov. J. Low
Temp. Phys. \textbf{9}, 22 (1983)].
\bibitem{14}F. Gompf and W. Reichardt,  (\href{https://inis.iaea.org/search/search.aspx?orig_q=RN:7253370}{Progress Report of the Teilinstitut Nucleare
Festk$\ddot{\text{o}}$rperphysik}, Fachinformationzentrum, Karlsruhe, 1975), p. 33.
\end{thebibliography}
\end{document}